\documentclass[fleqn,usenatbib]{mnras}

\usepackage{newtxtext,newtxmath}
\usepackage{multirow}

\usepackage[T1]{fontenc}

\DeclareRobustCommand{\VAN}[3]{#2}
\let\VANthebibliography\thebibliography
\def\thebibliography{\DeclareRobustCommand{\VAN}[3]{##3}\VANthebibliography}

\usepackage{graphicx}	
\usepackage{amsmath}	

\title[Antenna Structure and Orientation in 21\,cm Cosmology]{Impact of Antenna Structure and Orientation on Forward-Modelled Global 21 cm Signal Recovery}

\author[J.H.N. Pattison et al.]{Joe H. N. Pattison,$^{1,2}$\thanks{E-mail: jhnp2@cam.ac.uk}
John M. Cumner,$^{1,2}$\thanks{E-mail jmc227@cam.ac.uk}
Dominic J. Anstey,$^{1,2}$
Saurabh Pegwal,$^{3}$
Wessel Croukamp,$^{3}$
\newauthor
Dirk I. L. de Villiers,$^{3}$
and Eloy de Lera Acedo$^{1,2}$
\\
$^{1}$Astrophysics Group, Cavendish Laboratory, J.J. Thomson Avenue, Cambridge, CB3 0HE, UK\\
$^{2}$Kavli Institute for Cosmology, Madingley Road, Cambridge, CB3 0HA, UK\\
$^{3}$Department of Electrical and Electronic Engineering, Stellenbosch University, Stellenbosch, 7602, South Africa.
}

\date{Accepted XXX. Received YYY; in original form ZZZ}

\pubyear{\the\year{}}

\begin{document}
\label{firstpage}
\pagerange{\pageref{firstpage}--\pageref{lastpage}}
\maketitle

\begin{abstract}
The redshifted 21 cm absorption trough from cosmic atomic hydrogen is one of the most promising probes of the early Universe, but its detection is challenged by bright foregrounds and instrumental systematics. In this work we quantify the impact of antenna mismodelling on signal recovery within a fully Bayesian, forward-modelled data analysis pipeline. We show that discrepancies between simulated and modelled antenna beams lead to frequency-dependent errors in antenna temperature that can bias parameter inference. In particular, we demonstrate that orientation mismatches at the level of $\sim 0.25^\circ$ can significantly bias recovered signal parameters in typical observing scenarios. However, we also show that Bayesian evidence can be used to infer antenna orientation within this precision by scanning over model realisations. For structural mismodelling, we find that broadband recovery of all signal parameters requires accurate beam knowledge, but that partial recovery remains possible. In particular, signal frequency and width can be robustly recovered under restricted frequency bands even when the antenna structure is imperfectly modelled, while signal depth remains highly sensitive to beam errors. These results quantify the level of beam knowledge required for forward-modelled global 21 cm experiments and highlight the importance of observing strategy and antenna design in mitigating beam–sky coupling systematics.
\end{abstract}




\section{Introduction}
There is a 5.88\,\(\mu\)eV (21\,cm wavelength) energy gap between the parallel and anti-parallel spin states of neutral hydrogen.
The absorption and emission of energy allowing this spin flip to take place in the early Universe may well be one of the most promising tools to characterise the Dark Ages and Epoch of Reionisation.
This redshifted `21\,cm' signal from cosmic neutral hydrogen is a subject of ongoing inquest as astrophysicists try to understand what was happening between the emission of the Cosmic Microwave Background (CMB) and the reionisation of the Universe.
Finding this signal has proved to be a challenging task, as this signal is far dimmer than the galactic foregrounds it sits behind.
As of yet the only claimed detection of this signal comes from the Experiment to Detect the Global Epoch of Reionization Signature (EDGES) \citep{Bowman2018AnSpectrum}.
The EDGES team found a flattened Gaussian signal with a depth of 500\,mK has sparked debate within the community, asking if this signal was due to new, as of yet unexplained physics.

Explanations for the depth and shape of the signal have included millicharged dark matter causing a more rapid cooling of the early Universe than was anticipated by standard models \citep{Liu2019RevivingCosmology}, an excess radio background causing hydrogen to have a much lower relative temperature to background than expected \citep{Fialkov2019SignatureSpectrum, Mittal2022ImplicationsSurveys}, and competing sources of reheating \citep{Gessey-Jones2022ImpactSignal}.
However, there have been claims that this signal rose from unresolved systematics within the data \citep{Hills2018ConcernsData, Sims2019TestingSelection}.
As of yet, no follow up experiments have been able to find this signal \citep{Singh2022OnBackground}.

This signal, having an expected depth of no more than 500\,mK in the most extreme cases, is vulnerable to being masked by systematics when they are not correctly accounted for.
These systematics, ranging from the impact of chromaticity \citep{Anstey2020AExperiments, Cumner2024TheExperiment}, ionospheric effects \citep{Shen2021QuantifyingObservations}, effects from the horizon and far-field soil \citep{Bassett2021LostAnalysis, Pattison2024ModellingForegrounds}, the impacts of changing weather conditions and near-field soil \citep{Pattison2025GlobalConditions}, and unmodelled occultation of the sky \citep{Pattison2026QuantifyingCosmology}, could all result in a non- or biased detection of the signal.

To deal with these systematics there are two approaches, avoidance or modelling.
All experiments try to avoid these systematics to the best of their ability, with the Shaped Antenna measurement of the background RAdio Spectrum (SARAS) \citep{Singh2017SARASSignal} building an antenna that is nearly achromatic across its observing band to avoid chromaticity, EDGES \citep{Bowman2018AnSpectrum} being located on a flat plane to avoid the horizon, and `Probing Radio Intensity at high-Z from Marion' (PRI\(^{\mathrm{Z}}\)M) \citep{Philip2019ProbingInstrument} being located on Marion island to avoid radio frequency interference.

One of the experiments trying to uncover this global, or sky-averaged, 21\,cm signal is the Radio Experiment for the Analysis of Cosmic Hydrogen (REACH) \citep{deLeraAcedo2022The7.528}.
The philosophy of the REACH experiment is that total systematic avoidance is impossible, so modelling for systematics is a necessity.
To this end the first phase of the REACH experiment is an all-sky observation with a dipole radiometer in the Karoo radio reserve South Africa, the data from which will be analysed with a fully Bayesian forward modelled pipeline \citep{Anstey2020AExperiments}.

The first phase of REACH uses a hexagonal bladed dipole sitting atop a serrated ground plane \citep{Cumner2022RadioCase}.
The structure of the raised ground-plane, and placement of the antenna, while simple and affordable in principle, are challenging to build to the high accuracy required in practice science concerning the epoch of reionisation.
With these initial difficulties, and cycles of heating and cooling in the Karoo semi-desert causing the ground-plane to become warped over time, the true dimensions of the REACH instrument vary slightly from those initially envisioned.
In practice any experiment will see slight differences between design and realised physical structure, and any ground plane exposed to changing environmental conditions will see structural evolution over time.

It has been shown that incorrect assumptions on the gain pattern of an antenna to the level of one part in ten thousand \citep{Cumner2024TheExperiment}, may lead to a masking of the signal behind systematics.

While polynomial foreground models may appear more robust to beam uncertainties, this robustness arises from their ability to absorb instrumental systematics into flexible functional forms.
This can, however, lead to degeneracies between foregrounds, systematics, and the cosmological signal, potentially resulting in biased or non-physical inferences. 
In contrast, physically motivated forward-modelling approaches impose stronger requirements on instrument knowledge, but offer improved interpretability and the ability to explicitly identify and constrain systematic effects.
This work aims to quantify these requirements for antenna beam modelling within such a framework.

Section \ref{sec:antennamethods} discusses the methodology used to measure the REACH instrument, the Computational Electromagnetic (CEM) antenna models used in this work, and the data analysis process used.
Section \ref{sec:antennaresults} discusses the impact of incorrect modelling of antenna orientation and structure on the recovery of the global redshifted 21\,cm signal.
In Section \ref{sec:antennaconclusions} we summarize the impact of physical complexity on a global 21\,cm measurement using the REACH methodology.

\section{Methods}
\label{sec:antennamethods} 
In this Section we outline the methods by which we used to complete the analysis presented in this work.
We detail the means by which the REACH radiometer dimensions were measured in Section \ref{sec:antennameasurement}, and how we use this to build the gain patterns used in Section \ref{sec:antennamodelling}.
In Section \ref{sec:antennaanalysis} we briefly outline the REACH data analysis pipeline.

\subsection{Antenna Measurement}

\begin{figure*}
    \centering
    \includegraphics[width=\linewidth]{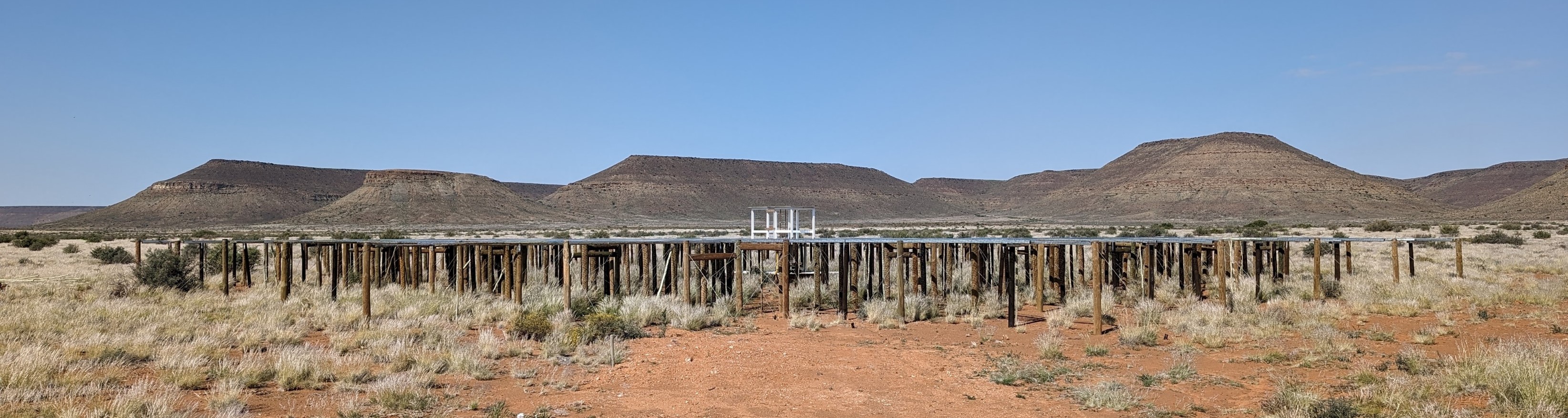}
    \caption{Side profile of the REACH radiometer taken from the north west side of the instrument. The image shows the REACH serrated ground plane elevated 1\,m above the soil by a series of wooden posts. On top of the metal ground plane in the centre of the image we see a table, painted white, supporting the REACH hexagonal dipole blades. Image courtesy of Rohan Patel.}
    \label{fig:reachside}
\end{figure*}

The REACH radiometer consists of a hexagonal bladed dipole sitting atop a raised mesh ground plane.
The ground plane is roughly aligned with the Earth's north-south axis and consists approximately of a \(20 \times 20\)\,m square with each side having 5 triangular serrations of length 6\,m and width 4\,m.
This ground plane was designed to be held 1\,m above the Karoo soil by wooden posts and metal u- and c- channels.
The dipole sits near to the centre of the central square with its y axis aligned with the ground plane itself.
A side profile of the ground plane and dipole is shown in Figure \ref{fig:reachside}.

\label{sec:antennameasurement}
\subsubsection{Orientation}
The orientation of the dipole was measured with a traditional magnetic compass, placing said compass flush against the top of the dipole blades, parallel to the centerline of the aluminium blades.
The dipole blades were measured to face \(21^\circ\pm 1.5^\circ\) east of magnetic north.
This value, measured on the 15\(^{\mathrm{th}}\) of August 2025, was then modified with magnetic declinations found from the 2025 World Magnetic Model \citep{Chulliat2025The2025-2030}, to give an orientation \(3.5^\circ\) west relative to true north.
The table on which the dipole blades sit was found to have an orientation of \(0.5^\circ\) east of true north.
The orientation of the ground plane was measured from the centre wooden post on the north side, such that the steel mesh ground plane would not corrupt measurements.
This was shown to be facing \(3^\circ\) east of true north.

The error in the absolute value of these measurements relative to true north sit between \(0.5\)--\(1.5^\circ\), which arise from the precision of the compass, the accuracy of measurement in the presence of ferrous materials, as well as the conversion from magnetic to true north from South Africa using the World Magnetic Model -- the conversion between magnetic and celestial north in this region has an error of \(\pm0.4\)--\(0.5^\circ\) \citep{Chulliat2025The2025-2030}.
However, the relative angles of the dipole blades, dipole table and ground plane as well as the angle of the ground plane relative to magnetic north are all independently corroborated through the ground plane topography measurements detailed later in this section.

\subsubsection{Measurements of Radiometer Morphology}
All dimensions of the dipole on top of the ground plane were measured with tape measure.
The dipole blades were found to be within design specification to sub-millimetre levels, as a result we will not focus on this in our subsequent analysis. There was found to be a small rotational offset between the wooden table which the dipole sits upon and the blades, measured to be $5\pm 3\text{~cm}$ at one end, leading to a $3^\circ\pm1.5^\circ$ rotation to the west. As the locations of the dipole legs can be obtained while measuring the ground plane topography this allows for an additional measurement of the angle of the dipole blades.

The measurement of the level of a raised ground plane is a non-trivial task, here we detail the steps taken to measure the topography of the ground plane such that an improved model of it may be built in CST Studio Suite\textsuperscript{\textregistered} compared to the idealised flat model. This process was aimed to be simple to execute by a small team with limited equipment and minimal expense, such that the severity of the problem can be assessed; and then further more detailed or high tech measurements taken at a later date if required.

Firstly, measurements across all serrations were taken, from base to tip of each serration across their diagonal, as well as a measurement across the base of each.
This allows us to build an outline of the ground plane, which we anticipate being in part different to the idealised model.
Following this we aimed to build a model of the topography of the surface of the ground plane.
To do this we establish four points around the radiometer, approximately 10~m away from the ground plane and centrally positioned on each side of the antenna.
At each of these points we placed a dumpy level, and measured the angular difference between two sets of poles used to hold up the ground plane.
Comparing these angular measurements to the measured physical distances between these poles allows for the triangulation of each point the dumpy level was placed relative to a point half way between the poles as per 
\begin{equation}
    D = \frac{d}{2\tan(\frac{1}{2}(\theta_1-\theta_2))},
\end{equation}
\noindent where the distance from the dumpy level to a given point is designated \(D\), the physical distance between two measured points is termed \(d\), and the angular separation between the poles is given by \(\theta_1-\theta_2\).

\begin{figure}
    \centering
    \rotatebox{-90}{\includegraphics[width=\linewidth]{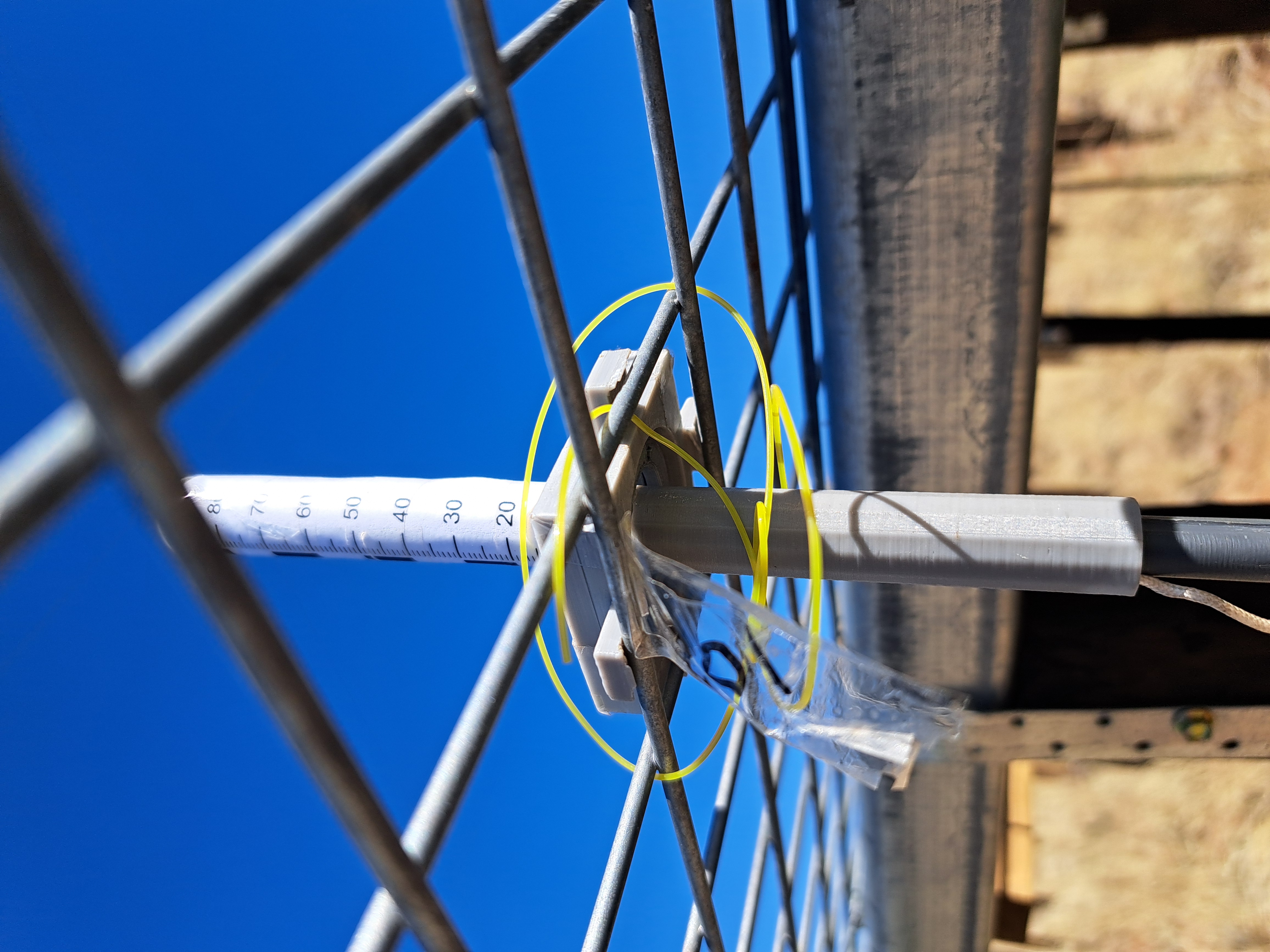}}
    \caption{Measuring device used to determine the topography of the ground plane of the REACH instrument. Location of implement is marked with yellow, non-conductive twine, and numbered.}
    \label{fig:measurestick}
\end{figure}

To allow measurement atop the ground plane a measuring implement was fashioned - resembling a ruler sitting in a cradle attached to a plumb bob (see Figure \ref{fig:measurestick}).
This device is held vertical by the plumb bob, allowing the dumpy level to be rotated to any point this device was placed, and thereby read off where on the ruler it found level.
From this, we are able to ascertain the relative heights of any point in the ground plane.
This device is placed at a number of positions across the ground plane, all of which were marked with yellow, non-conductive, twine and labelled with a number to allow for repeated measurement.
These locations were chosen such that they would somewhat evenly cover the span of the surface, as well as particular spots that appeared to be more deformed than others.

From each dumpy level position we measure all marked points in the half of the ground plane nearest the level.
For each measurement, we note the relative angle of the dumpy level to magnetic north, to $\pm0.25^\circ$, and the height on the devices ruler, to $\pm1\text{~mm}$. With this measurement repeated taken from two different places it is then possible to triangulate each device location and verify the height measurement.

These measurements were taken first on the north and south sides of the REACH antenna, before taking east and west measurements.
To calibrate the height of the dumpy level for the east and west measurements, the level was adjusted such that a measurement of two previously measured points at the north and south sides matched to within 1\,mm.
Concordance of height measurements was reached when measurements from two locations agreed to within 3\,mm.
To reach concordance for all points we required the re-measurement of height of the ground plane at three different ground plane positions, indicating a robust measurement procedure with a first time success rate of 98.2\%.

This process allows the calculation of the relative deformation of the ground plane with respect to other measured points.
These were then calibrated relative to the antenna blades by measuring the vertical distance from the antenna blade to a point on the ground plane near the antenna, and the value of the height on the device at the same location on the ground plane.

\begin{figure}
    \centering
    \includegraphics[width=\linewidth]{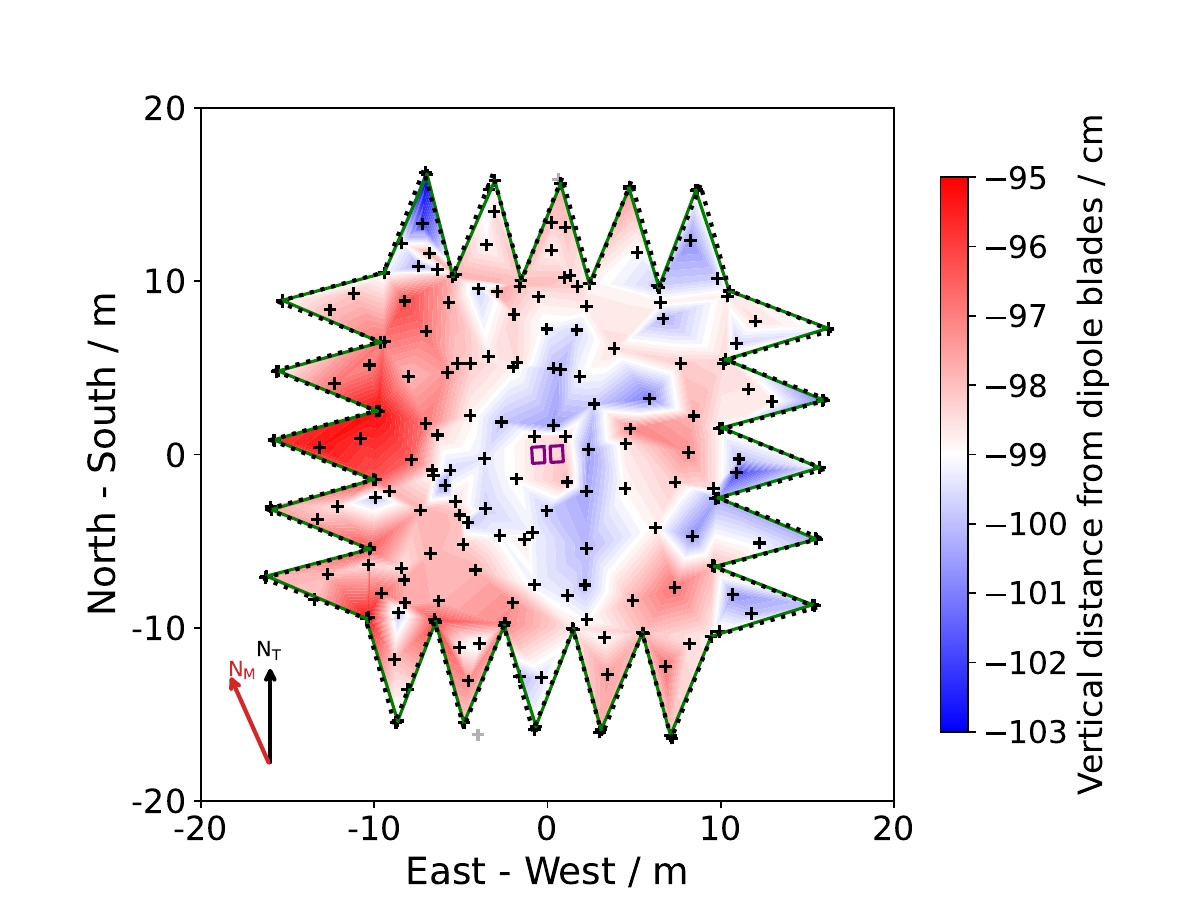}

    \caption{Topographic map of the ground plane of the REACH antenna. The black dotted lines indicate the dimensions of the ground plane as designed, the solid green lines the measured dimensions of the extent of the ground plane. The black crosses indicate measurement locations on the ground plane, the grey crosses represent measurements which were outside the ground plane bounds following fitting and discounted. The table upon which the dipole sits is denoted in purple. True north is located along the y-axis, magnetic north is indicated by the red arrow offset by $24^\circ$.}
    \label{fig:groundplanetopog}
\end{figure}

These results allowed us to produce a topographic map of the REACH ground plane, as seen in Figure \ref{fig:groundplanetopog}.

Using a direct calculation of the level positions resulted in many of the measured device locations being outside the expected footprint of the ground plane. To correct for this, the location and offset angle to magnetic north for each level position was allowed to vary and fit for under the condition that the measured points should fit inside the ground plane, particularly those known to be at the turning points of the ground plane were required to sit close to the expected outline. 
This greatly improved the results, although some points remained outside the footprint - these were discounted when constructing CAD models.


From Figure \ref{fig:groundplanetopog} we are able to see that the highest the ground plane peaks is 95.2\,cm below the dipole blades, with its lowest point being 103.0\,cm beneath.
These heights have a standard deviation of 1.27\,cm, with the extremes in heights coming from the serrations.

\subsection{Antenna Modelling}
\label{sec:antennamodelling}
From measurements taken on site, as well as the antenna design specifications we build a series of CAD models for simulation within CST Microwave Studio\textsuperscript{\textregistered}.
Our antenna models exist in four layers of complexity, based on the variation of three different aspects of the antenna structure as measured:
The flatness of the ground plane, the structure of the ground plane, and the angle of the dipole atop the ground plane.

The base level of complexity is an antenna built to initial specification; a flat \(20 \times 20\)\,m ground plane, each face of the ground plane having 5 triangular serrations of length 6\,m and width 4\,m.
That is to say an antenna that has a Flat ground plane (no warping has taken place due to weight or weathering), with the Ideal ground plane dimensions, as specified in \citet{Cumner2022RadioCase},
and dipole blades that have their axis aligned Straight relative to the ground plane - we designate this beam FIS.

\begin{figure*}
    \centering
    \includegraphics[width=\linewidth]{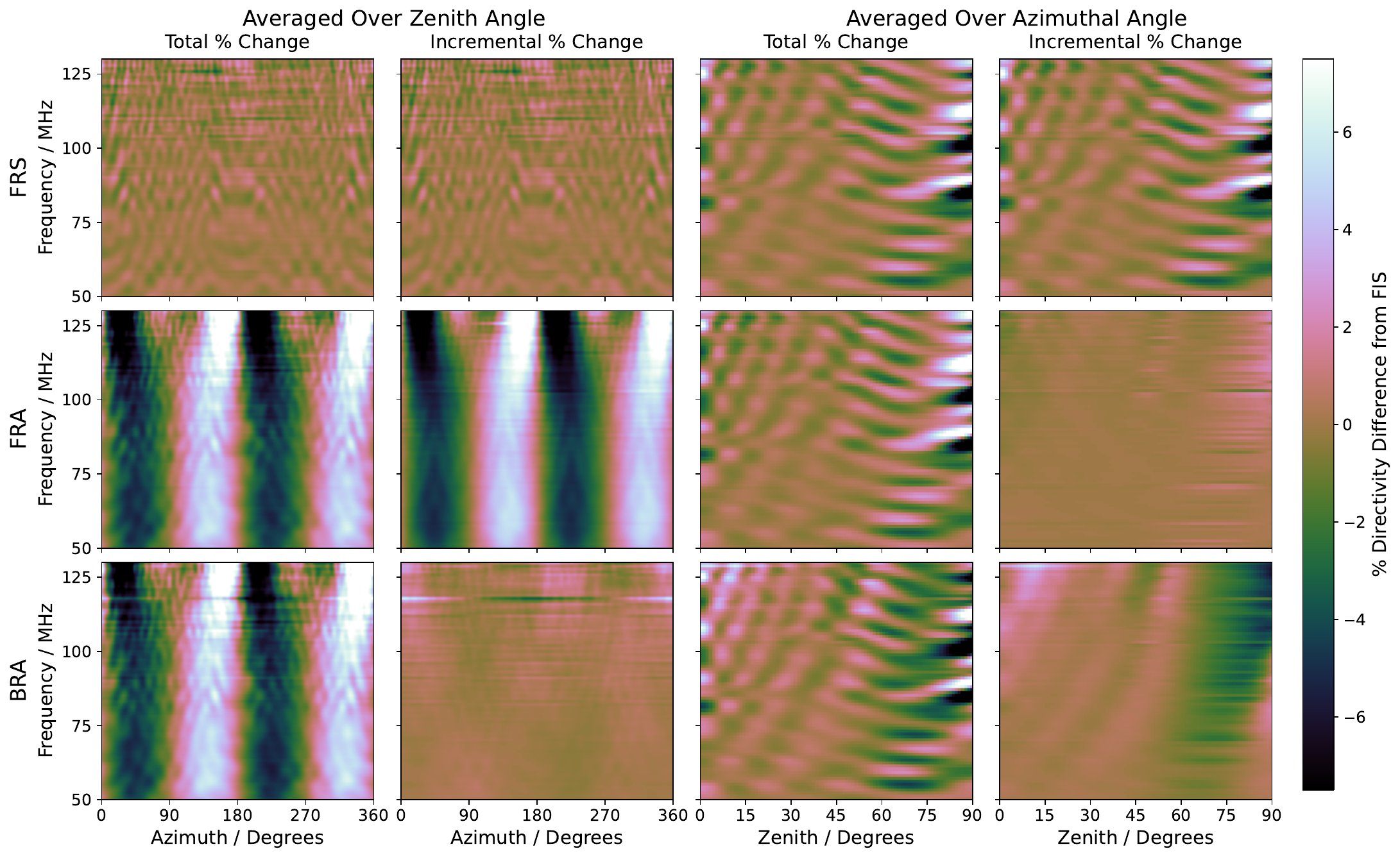}
    \caption{Percentage change in the directivity of the REACH dipole antenna as complexity of the modelled antenna increases. Each panel shows a waterfall plot of antenna chromaticity. The `cubehelix' \citep{Green2011AImages} colourmap giving the mean percentage directivity difference per unit solid angle, for the two left-hand columns this is for a given azimuthal angle, averaged across all zenith angle, and for the two right-handed columns this shows directivity at a given angle from zenith for all values of azimuthal angle. The left hand column shows the cumulative change in directivity from an idealised antenna model (FIS) as the complexity of the modelled antenna increases, and the right hand column indicates the incremental percentage change in directivity with each layer of additional complexity. The ideal antenna has a ground plane built to originally specified dimensions, with a dipole angled such that the y axis are is aligned to the north-south axis, and the ground plane is perfectly flat. The top row changes the ground plane structure from having the idealised dimensions with perfect serrations, compared to those measured on site (FRS). The second row rotates the dipole relative to the ground plane to match measured values (FRA). The bottom row adapts the topography of the ground plane to match that measured on site (BRA).}
    \label{fig:changingchrom}
\end{figure*}

Our second layer of complexity changes our ground plane serrations such that they are Realistically shaped (as measured), leaving all other aspects the same, we designate this beam FRS.
Next, we design a model with a dipole that has been appropriately Angled on the ground plane, with all other aspects the same as the FRS beam, this beam will be called FRA.
Finally, we change our ground plane topography so that we include the Bumps in the mesh, this beam is designated BRA.
The difference in gain pattern, as computed by CST, of these antennas is shown in Figure \ref{fig:changingchrom}.

We see from Figure \ref{fig:changingchrom} that 
the rotation of the dipole on the ground plane has very little impact on beam structure when averaged over zenith angle, but that it can increase or decrease gain across azimuthal angle by upwards of 7\%.
We also see that variation in the structure of the ground plane, moving from FIS to FRS produces ripples across zenith and azimuthal angle.
This is also true when we introduce variations in the topography of the ground plane, seeing a change in gain across zenith angles as our gain pattern changes from FRA to BRA.
The scale of these ripples are consistent across azimuthal angle, but increase as zenith angle and frequency increases.
This increase in gain pattern difference at high frequencies is unsurprising, as the peak gain of our dipole pattern moves further from zenith as frequency increases \citep{Cumner2022RadioCase, Pattison2026QuantifyingCosmology}.

\subsection{Data Analysis}
\label{sec:antennaanalysis}
To analyse our results we use the data generation, and recovery arms of the REACH fully Bayesian, forward-modelled, data analysis pipeline \citep{Anstey2020AExperiments}.

The data generation arm of the REACH pipeline relies on the Global Sky Model (GSM) \citep{deOliveira-Costa2008AGHz}, a model that uses principle component analysis to decompose diffuse galactic spectral variation into dominant and subdominant modes, and then uses the three dominant modes to generate a map of diffuse galactic emission at a given frequency.
We utilise this model to generate a map of the spectral indices across the sky.
We do this by generating two maps at different frequencies and calculating the required spectral index to map one of these onto the other.

The spectral indices at each pixel will be used to scale one of the generated base maps to give the brightness temperature per pixel per unit frequency across the working frequency band of our antenna.
Once these maps have been created and scaled, they are rotated according to time, date, and location of simulated observation, and masked with a reflective and emissive horizon \citep{Pattison2024ModellingForegrounds}.
To this we inject a 21\,cm signal of known parameters, some expected instrumental noise, and the power arising from the CMB.
These maps are then multiplied by the gain pattern of an antenna, and the power per frequency of the beam-weighted sky map is integrated across all angles to give a simulated antenna temperature per unit frequency.
For this analysis we assume perfect absolute calibration of our radiometer, though in practice this is a more complicated task \citep{Kirkham2025AccountingExperiments, Leeney2025RadiometerLearning}.

To uncover the 21\,cm signal, we turn to the analysis arm of the REACH pipeline.
This divides the sky into a series of coarse-grained regions based on the spectral index map we derive from the GSM, such that the intra-region pixels each have a similar spectral index.
This map is once again rotated, and masked by a horizon.
An integrated all-sky product of this map and the gain of a simulated antenna is taken to be our foreground model, the exact power of which will be dependent on the spectral indices of each region.
To determine parameter values we design a likelihood that tries to minimise the difference between the antenna temperature of our dataset, and an antenna temperature derived from our foreground model plus some model of the 21\,cm signal.

Utilising the \textsc{polychord} \citep{Handley2015PolyChord:Sampling, Handley2015Polychord:Cosmology.}  nested sampling algorithm to fit this likelihood we recover the highest likelihood parameters as well as model evidence, which we can use to compare the Bayesian odds of one model over another.
This algorithm generates a random series of points drawn from the priors, known as `live points', and then iteratively discards the lowest likelihood point before generating another with a higher likelihood.
It does this until some stopping criterion has been met, shrinking the multidimensional posterior space around the highest likelihood peak.

For a full description of the REACH data analysis pipeline, see \citet{Anstey2020AExperiments,Anstey2023UseModelling, Pattison2024ModellingForegrounds, Pattison2025GlobalConditions}

\section{Results}
\label{sec:antennaresults}
In this Section we outline the results of our investigation.
Section \ref{sec:oriresults} discusses the antenna temperature difference observed when our entire radiometer structure is rotated, and the ability of the REACH pipeline to fit the 21\,cm signal with an unmodelled rotation.
Section \ref{sec:structureresults} discusses the impact of a mismodelled antenna structure on the ability of a physically motivated forward model to find the 21\,cm signal.

\subsection{Orientation}

\label{sec:oriresults}
To investigate the impact of assumed versus correct antenna orientations on the recovery of the 21\,cm signal we take an idealised antenna (FIS), and perform a number of simulated observations.
\begin{figure*}
    \centering
    \includegraphics[width=\linewidth]{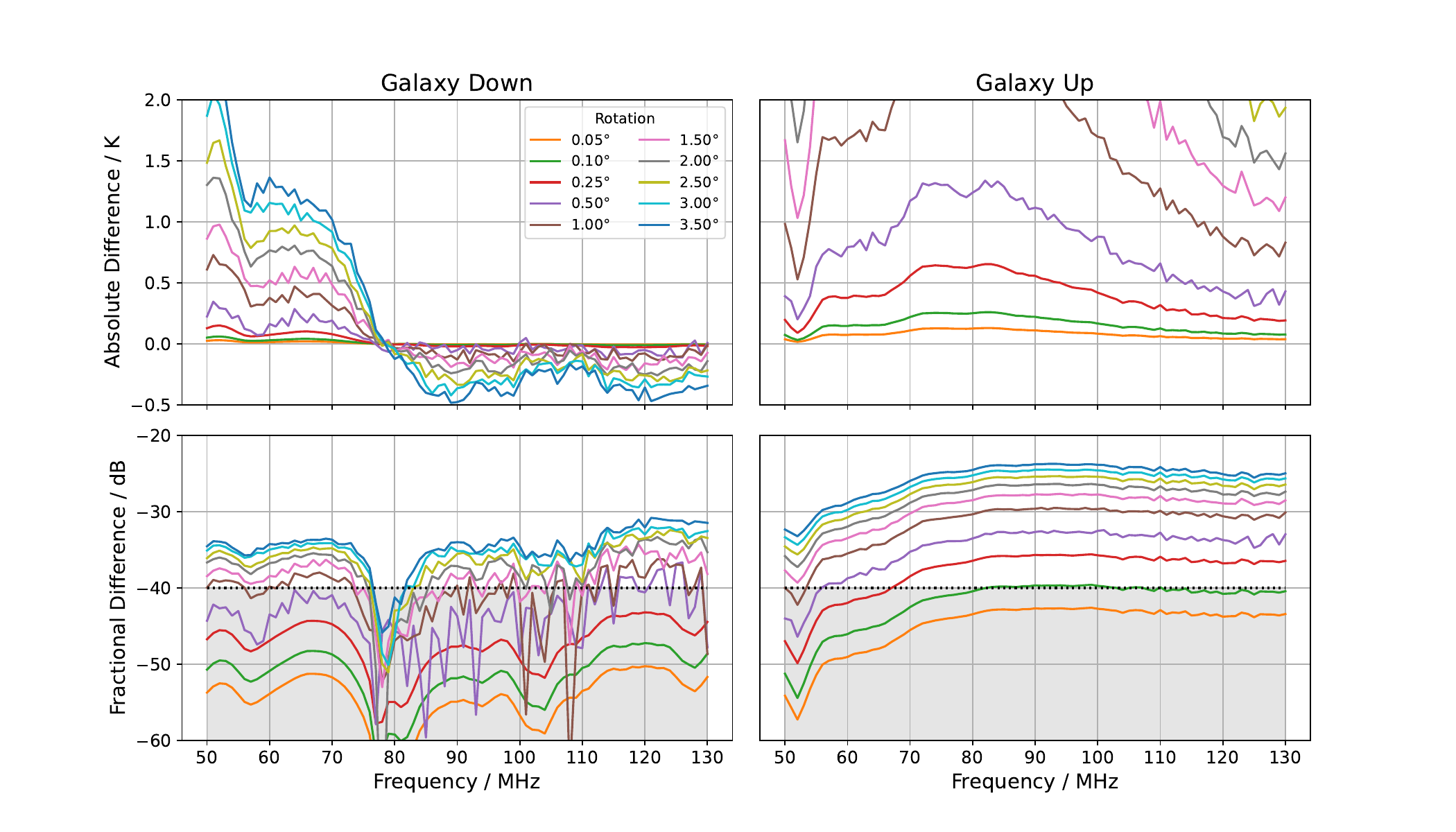}
    \caption{Absolute and fractional difference of antenna temperature per unit frequency when there is an angular mismatch between the antenna observing the sky in the simulation and model. Each coloured line refers to a different angular mismatch in degrees, as shown by the legend. The gain of the antenna beam is discretely sampled at 1\,MHz intervals.  The dotted black line and grey shaded region indicates a fractional difference of 100 parts per million. Each set of data represents the average antenna temperature over the course of three hours. The left hand side plots represent a `Galaxy Down' case in which our simulation begins 2019-10-01 00:00:00, during this period the galactic centre is near or below the horizon. The right hand side plots represent a `Galaxy Up' case in which our simulation begins 2019-07-01 00:00:00, during this period the galactic centre is above the horizon.}
    \label{fig:galupdowndata}
\end{figure*}

These observations will be divided into two categories.
The first we will refer to as an `Galaxy Up' observation, mirroring \citet{Anstey2020AExperiments} - this observation begins 2019-07-01 00:00:00 UTC (LST = 16.5 hour angle) and describes a time where at the Karoo Radio Reserve the galactic centre is above the horizon.
The second will be referred to as `Galaxy Down', beginning 2019-10-01 00:00:00 UTC (LST = 22.5 hour angle) and describes a time where the galactic centre will be at or below the horizon.

In our first test, we compare the antenna temperature per unit frequency for observations where we simulate a rotation of the antenna and ground plane by a given number of degrees westward.
In Figure \ref{fig:galupdowndata} we show the difference in antenna temperature between observations of rotated antennas.

We see from Figure \ref{fig:galupdowndata} that when the galaxy is up, a small rotation will quickly lead to large temperature mismatches.
A rotation of the radiometer in excess of \(1^\circ\) will create a consistent temperature increase in excess of 1000 parts per million across frequency.
Even a \(0.25^\circ\) rotation of antenna and ground plane will lead to difference in antenna temperature of upwards of 300 parts per million.
This is a level of error in antenna temperature which, when inhomogeneous across frequency, has been shown to bias or obfuscate a simulated recovery of the Global 21\,cm signal using a physically motivated model \citep{Cumner2024TheExperiment, Pattison2026QuantifyingCosmology}.
In the Galaxy Down case, as shown in Figure \ref{fig:galupdowndata}, a rotation of the instrument is more forgiving.
A rotation of the instrument of one degree or below will give a fractional difference of antenna temperature less than 100 parts per million.

\begin{figure*}
    \centering
    \includegraphics[width=\linewidth]{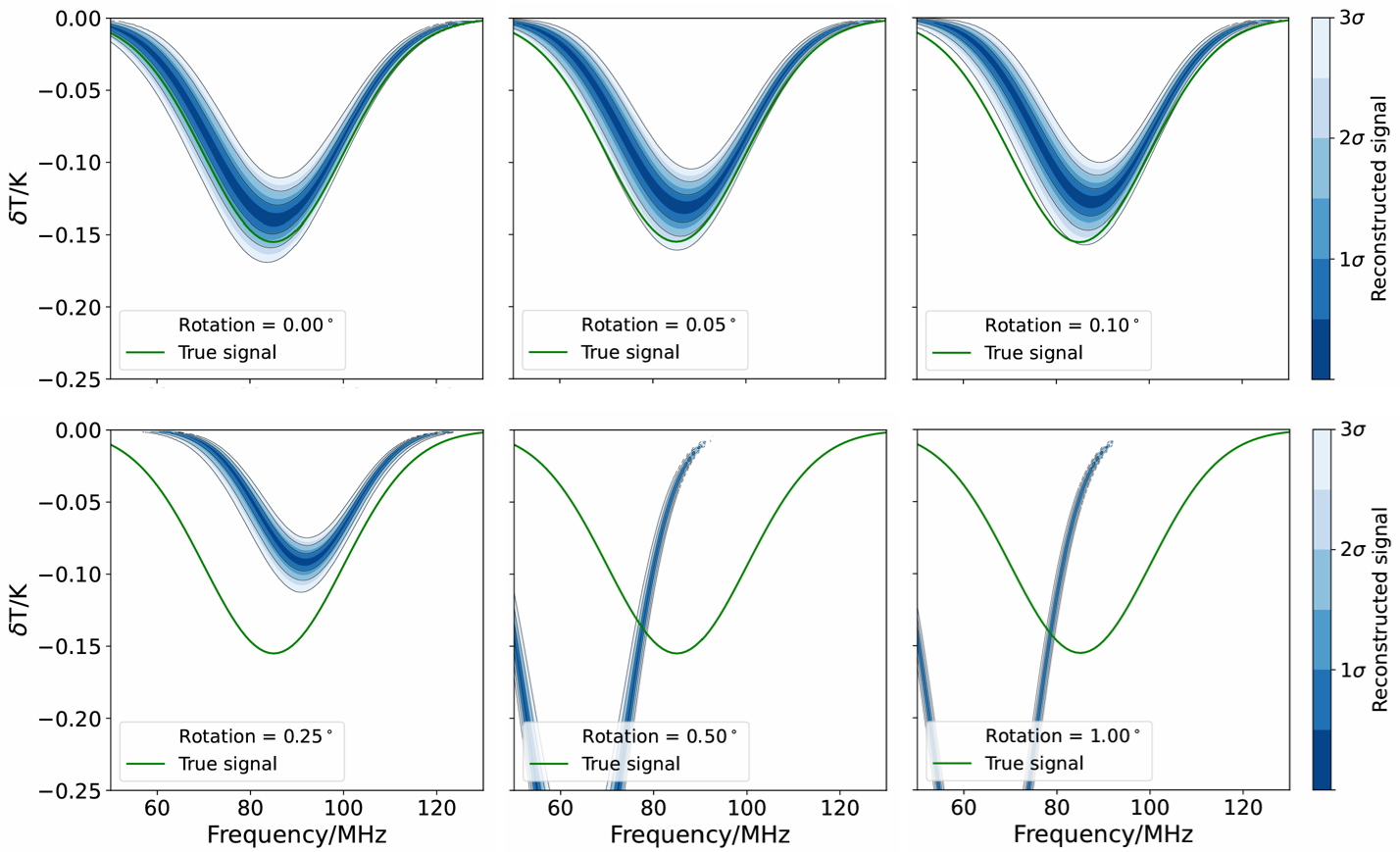}
    \caption{Simulated recovery of a redshifted 21\,cm signal with shifting antenna orientations. The injected signal is a Gaussian with a central frequency of 85\,MHz, a width of 10\,MHz and a depth of 0.155\,K.  The shaded blue areas represent credible intervals of signal recovery, and the green line shows the injected signal. The simulated observation begins 2019-10-01 and lasts for three hours, during this period the galaxy remains below or near the horizon. The recoveries use a beam weighted sky-model in the REACH pipeline where there is a \(0.00^\circ\) mismatch between the antenna used to generate the data, and that used in our recovery model (top left), a \(0.05^\circ\) mismatch (top middle), a \(0.10^\circ\) mismatch (top right), a \(0.25^\circ\) mismatch (bottom left), a \(0.50^\circ\) degree mismatch (bottom middle), and a \(1.00^\circ\) degree mismatch (bottom right).}
    \label{fig:orisignalrecovery}
\end{figure*}

A level of error at or below 100 parts per million should allow for signal recovery.
We therefore investigate the Galaxy Down case further.
In Figure \ref{fig:orisignalrecovery} we fit a set of Galaxy Down data with six models, each of which assumes a different orientation of the antenna and ground plane with respect to the sky.
We fit for these datasets with a 40 sky region, time separated model, with 25 live points per region.
Our model has priors on the signal with a central frequency between 50--200\,MHz, a width between 10--20\,MHz, and a depth between 0--400\,mK.
The first of these models acts as a control, where we correctly match our modelled orientation with that which we used to create the data, and subsequent models where these are mismatched.
We see that as little as an unmodelled \(0.25^\circ\) degree  rotation will bias recovery of the injected signal beyond \(3\sigma\), with any rotation beyond this saturating the model priors; and thereby drives the inferred parameters to the boundaries of the prior, preventing reliable parameter inference. This implies that even in scenarios where the galaxy is at or below the horizon, antenna orientation must be modelled to a precision below \(0.25^\circ\) to enable unbiased recovery of signal parameters within a broadband forward-modelling framework. This level of precision is beyond what is possible with a magnetic compass.

We note that this requirement is not expected to be universal, but instead reflects the coupling between the antenna beam and the anisotropic sky brightness distribution. In particular, the sensitivity to orientation errors will depend on the degree of beam asymmetry and the position of bright foreground structures relative to regions of high gain. Quantifying how this requirement scales with antenna design and observing strategy is an important direction for future work.

\begin{figure}
    \centering
    \includegraphics[width=\linewidth]{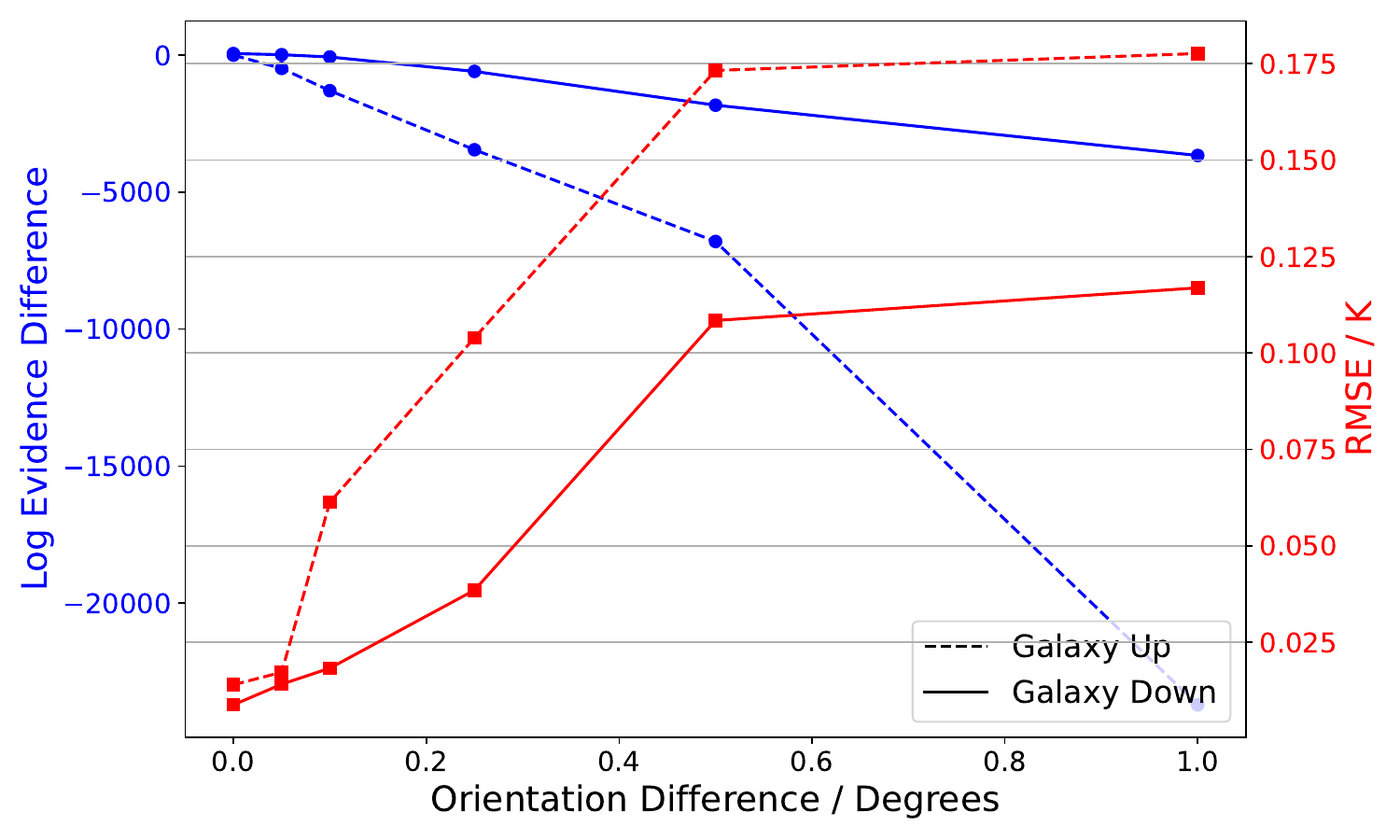}
    \caption{Changes in log evidence and RMSE of signal recovery as the mismatch in orientation between the antenna used to generate data, and that used to recover data changes. The blue lines indicate the difference in log evidence between the model plotted and a model in which the orientation in the data generation and signal recovery align perfectly. The red line indicates the RMSE of signal recovery. The dashed lines refer to the Galaxy Up case, the solid lines show the Galaxy Down case. }
    \label{fig:logevrmse}
\end{figure}

We see this further with Figure \ref{fig:logevrmse}, where we plot the log evidence and RMSE of signal recoveries with recovery models with changing orientations.
Both the Galaxy Up and Galaxy Down recoveries saturate the priors beyond a rotation of \(0.5^\circ\), settling at different locations in the prior space.
We do, however, note that the log evidence of our models decrease as the angular mismatch between our simulated data and our recovery model increases.
This implies that one could use the Bayesian evidence as a metric to fit for the overall orientation of the radiometer.
In the current structure of the REACH pipeline, we pre-calculate parameter independent `chromaticity functions' outside of the likelihood we draw from in our fitting to allow for the expedition of the process itself \citep{Anstey2020AExperiments}.
This pre-calculation is necessary for the pipeline in its current form to run within a practical timescale.
The orientation of the antenna would alter these chromaticity functions, so fitting for this as a parameter within the pipeline would make this pre-calculation challenging.
With a sufficient computational time, however, a number of fits could be run with a series of fine-grained antenna orientations within an expected set of error bounds, the highest evidence case of which would be taken to be the `true' orientation.
Using this fitting process we would be able to probe antenna orientation beneath the error floor of a physical magnetic model, and produce reliable, accurate signal recoveries, even when precise antenna orientation is not known a priori.

A full description of evidence and accuracy of signal recovery with changing model orientations is found in Table \ref{tab:orientation}.

\subsection{Antenna Structure}
\label{sec:structureresults}
To investigate the impact of an incorrectly modelled antenna structure upon the recovery of the 21\,cm signal, we compare the simulated antenna temperature per unit frequency when the sky is observed using our four antennas, as in Section \ref{sec:antennamodelling}.
These are: the antenna with a Flat, Ideally shaped ground plane, with a dipole who's y axis is Straight relative to said ground plane (FIS), the antenna that has a Flat, but Realistically shaped ground plane, with a Straight dipole (FRS), the antenna that has a Flat, Realistically shaped ground plane, with the dipole Angled on the ground plane as measured on site (FRA), and the antenna with a Bumpy, Realistically shaped ground plane with Angled dipole (BRA). 
These simulations begin on our Galaxy Up and Galaxy Down dates, and each last six hours. In this analysis, the BRA beam is treated as a proxy for a more realistic realisation of the instrument, incorporating measured structural complexity. It should not be interpreted as the exact ground truth response of the antenna, but rather as a representative model against which simpler descriptions can be compared.

\begin{figure*}
    \centering
    \includegraphics[width=\linewidth]{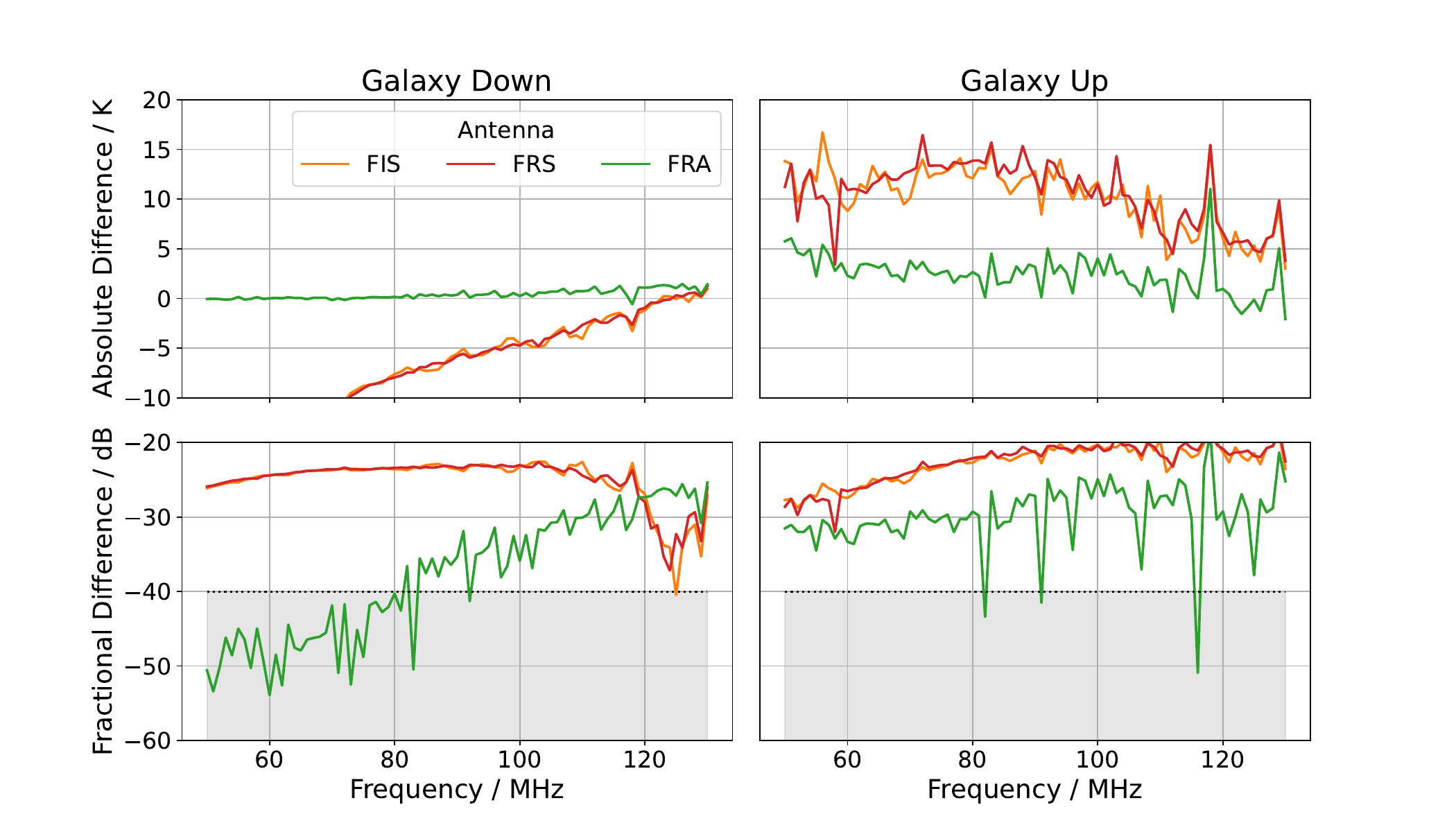}
    \caption{Absolute and percentage difference in antenna temperature between a simulated observation taken with the most complex beam (BRA), versus ones that reduce the complexity of the beam by subsequently removing defects in the ground plane (FRA) as shown in green, aligning the dipole with the ground plane (FRS) as shown in red, and fixing the ground plane to the ideal dimensions (FIS) as shown in orange. The gain of these beams are discretely sampled at 1\,MHz intervals.  The dotted black line and grey shaded region indicates a fractional difference of 100 parts per million. Each simulation lasts for six hours. The left hand side plots are our `Galaxy Down' case, beginning 2019-10-01 00:00:00, during this period the galactic centre is below or near the horizon. The right has side plots are our `Galaxy Up' case, beginning 2019-07-01 00:00:00, during this period the galactic centre is above the horizon.}
    \label{fig:galupdowndataant}
\end{figure*}

We see from Figure \ref{fig:galupdowndataant}, showing the absolute and fractional differences in antenna temperature observed with our different antenna models, that the Galaxy Up simulations observe a much larger difference in antenna temperature as we change model complexity with respect to the Galaxy Down case.
We also note from both sets of plots that the difference in antenna temperature immediately increases when we introduce a rotation to the dipole on the ground plane - across both dates the FIS and FRS antenna temperatures are sharply different to that of FRA and BRA.
In the Galaxy Up case, the difference between antenna temperatures for FRA and BRA simulations is of order 1000 parts per million, which is a difference an order of magnitude higher than would be required for an attempt at a recovery.
In the Galaxy Down case the difference in antenna temperature between BRA and FRA sits at order 100 parts per million up to 80\,MHz at which point it begins to rapidly rise.

The consequences of this are shown in Table \ref{tab:beams}.
Here we take two simulations, the six-hour Galaxy Down and Galaxy Up datasets generated with the BRA beam, and fit for them with a model using an antenna of each level of complexity.
We do this with a number of time-separated models, each with 45 sky regions, and 25 live points per region, using the same priors as in \ref{sec:oriresults}.

We observe that any level of beam complexity used in our model that differs from that used to create the data leads to saturation of the inferred parameters at the boundaries of the priors. In particular, the signal depth is driven to the limits of the prior in both the Galaxy Up and Down cases, while the signal width also becomes poorly constrained once the angle of the dipole on the ground plane is no longer accurately described. This indicates that incorrect beam modelling does not generically prevent signal recovery, but instead biases specific parameters and limits the ability to recover them reliably under broadband fitting.

\begin{figure}
    \centering
    \includegraphics[width=1\linewidth]{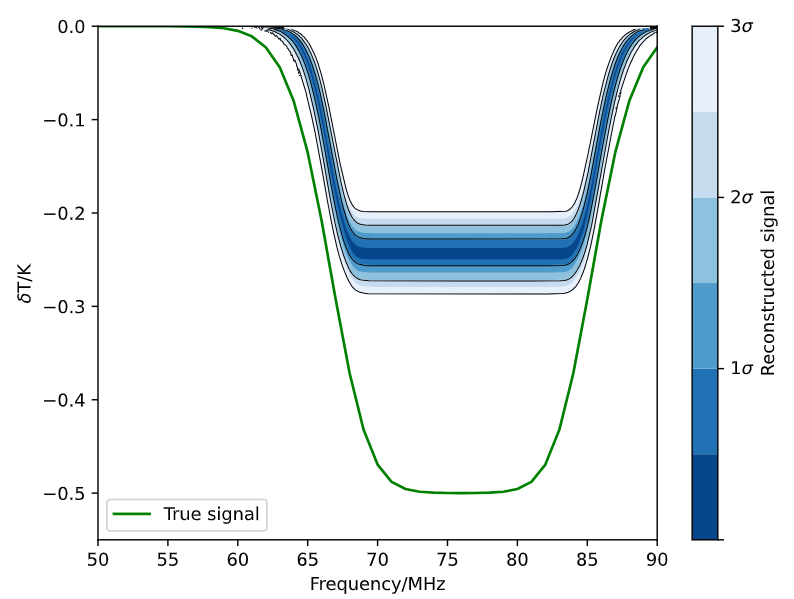}
    \caption{Simulated recovery of an EDGES-like redshifted 21\,cm signal. The inject signal is a flattened Gaussian with a centre frequency of 76\,MHz, a width of 19\,MHz, a depth of 500\,mK and a flattening factor of 7. The shaded blue areas represent credible intervals of signal recovery, and the green line shows the injected signal. The simulated observation begins 2019-10-01 and lasts for six hours. The data is generated with the BRA beam, and fitted with the FRA beam.}
    \label{fig:edgestest}
\end{figure}

However, as we noted, the Galaxy Down case appears to have somewhat consistent antenna temperatures between BRA and FRA up to 80\,MHz, at which point we break the 100 parts per million threshold.
To investigate the possibility of fitting a signal while inaccurately describing the topography of the ground plane we inject an EDGES-like flattened Gaussian with a depth of 500\,mK into our simulation with a BRA antenna and attempt to fit for it with the FRA model.
To do this we fit with a flattened Gaussian signal model with priors of: a centre frequency between 60--90\,MHz, a depth of 0--1000\,mK, a width of 1--40\,MHz, and a flattening factor of between 0--100.
From Figure \ref{fig:edgestest} we see that while we are able to accurately recover the centre frequency and width of the injected signal, the recovered signal depth is significantly biased. This demonstrates that different astrophysical parameters exhibit different sensitivities to beam modelling uncertainties. In particular, while the timing of key transitions in the early Universe (as traced by the signal frequency and width) may still be inferred under imperfect beam knowledge, the amplitude of the signal — which encodes information about the strength of heating and cooling processes — is much more sensitive to instrumental modelling. This highlights the importance of accurate beam characterisation for constraining the underlying astrophysics.

Being able to correctly locate this turning point in cosmic history without perfect antenna structure knowledge is encouraging, as will provide a window into the periods of transition between heating and cooling mechanisms, however it leaves us unable to describe strength of these cooling and heating sources in the early Universe - highlighting the importance of a comprehensive ground-plane model for recovering unbiased signal parameters, particularly the signal depth.

\textbf{Summary of recovery behaviour.}
Across both orientation and structural mismodelling cases, we find that the impact on signal recovery is not uniform across parameters.
While broadband recovery of all signal parameters requires an accurate description of the antenna beam, partial recovery remains possible under certain conditions.
In particular, the central frequency and width of the signal can be recovered even in the presence of moderate beam mismodelling, especially when restricting the analysis to frequency ranges where beam-induced spectral structure is minimised.
In contrast, the signal depth is significantly more sensitive to beam inaccuracies, and is typically biased or driven to prior boundaries when the beam model is incorrect.
This highlights that different astrophysical parameters place different requirements on beam knowledge.

\section{Conclusions}
\label{sec:antennaconclusions}

We show in this paper the effect of an inaccurately modelled antenna structure, and therefore gain pattern, on the recovery of a Global 21\,cm signal with a physically motivated foreground model.
Fully Bayesian, physically motivated forward modelling is the preferred analysis method of the REACH experiment, to allow the identification of systematics that may go unnoticed in a polynomial model.
Therefore, it is vital that we understand how these models are impacted by small uncertainties in the physical structure of the antenna.

To do this, we have used a selection of simulated versions of the REACH dipole antenna in conjunction with the REACH Bayesian data analysis pipeline to fit data generated with differing antenna simulations to that which are used within the analysis. 
We have presented results using antenna models including component rotations in addition to variations in the ground plane structure; these variations are motivated by measurements as detailed in Section \ref{sec:antennameasurement}.

Due to the convolution between sky power and the antenna gain pattern, we note that discrepancies within antenna temperature caused by imperfect modelling are increased by approximately 4 times when the galactic plane is in view of the antenna. 

We show that an orientation mismatch between simulated and modelled antenna beams of \(\geq 0.25^\circ\) may lead to biased signal recovery.
We demonstrate, however, that the Bayesian evidence of the recovered fit monotonically reduces as the discrepancy between orientation is reduced.
Therefore, once computational constraints are overcome, it would be possible to treat the antennas overall orientation as a model parameter to be fit for, in a similar manner as the sky's spectral index.

We demonstrate that variations within the ground plane structure of order centimetre generate order Kelvin differences in antenna temperature if not accounted for in the analysis antenna models. 
These discrepancies are seen to increase at higher frequencies, particularly above 100\,MHz, where the maximum gain of the antenna is directed closer to the horizon. 
Differences in the rotation of the antenna blades relative to the ground plane are seen to create antenna temperature differences larger than the Global 21\,cm signal at all frequencies, and it is therefore important to be accurately included within the antenna model used for data analysis in order to avoid biases in the recovered signal parameters, particularly in broadband analyses.
Despite the expected temperature difference; we show that, with a reduced frequency band, the REACH analysis pipeline is able to correctly identify the width and centre frequency of an EDGES-like flattened Gaussian signal, when ground plane discrepancies are not accounted for.
We also show that model evidence and goodness-of-fit increase as the gain pattern approaches the simulated ground-truth, reinforcing the possibility of using Bayesian evidence to fit for small imperfections in the beam model.

Overall, this work demonstrates that forward-modelled analyses of the global 21 cm signal impose stringent but quantifiable requirements on antenna beam knowledge. These requirements arise from the coupling between beam structure and the anisotropic sky, and depend on both antenna design and observing strategy. While imperfect beam modelling can bias certain parameters, particularly the signal depth, we show that partial recovery of key features such as signal frequency and width remains possible under appropriate conditions. These results provide guidance for both instrument design — favouring symmetric, zenith-oriented beams — and analysis strategies that minimise beam–sky coupling, and highlight the importance of developing methods to characterise and constrain beam uncertainties within forward-modelled frameworks.

\section*{Acknowledgements}

The first two authors, JHNP and JMC contributed equally to this work and should be considered joint first authors. Subsequent authors are listed in order that they provided assistance to measurements taken of the REACH instrument, and helped in the development of this work.

The authors thank Dennis Molloy for the design and production of the measuring implement used to ascertain the topography of the REACH antenna, Alessio Magro for his assistance on site, and Will Handley for his integral contributions to the REACH pipeline.

JHNP, JMC, DJA, and EdLA were supported by the Science and Technology Facilities Council.
We would also like to thank the Kavli Foundation for their support of REACH.

This research was also supported by the South African Radio Astronomy Observatory, which is a facility of the National Research Foundation, an agency of the Department of Science and Technology (Grant Number: 75322).

\section*{Data Availability}
The data that support the findings of this study are available from either of the first authors upon reasonable request.
\appendix
\section{Additional Tables}
\definecolor{fandango}{RGB}{181, 51, 137}
\definecolor{}{RGB}{0,0,0}
\begin{table*}
\centering
\begin{tabular}{lllllll}
\hline
& Rotation & F$_0$ (MHz) & Signal Width (MHz) & Depth (mK) & log($\mathcal{Z}$) & RMSE (mK) \\
\hline
\hline
\multirow{11}{*}{\rotatebox[origin=c]{90}{\textbf{Galaxy Down}}} 
&0.00$^\circ$ & 85.4 $\pm$ 0.8 & 14.3 $\pm$ 0.5 & 139.5 $\pm$ 9.5 & 6596.2 $\pm$ 0.9 & 8.8 $\pm$ 4.2\\
&0.05$^\circ$ & 86.8 $\pm$ 0.7 & 13.6 $\pm$ 0.5 & 132.0 $\pm$ 9.4 & 6555.4 $\pm$ 0.9 & 14.1 $\pm$ 4.7\\
&0.10$^\circ$ & 87.8 $\pm$ 0.7 & 12.9 $\pm$ 0.5 & 127.8 $\pm$ 9.3 & 6471.2 $\pm$ 0.5 & 18.3 $\pm$ 4.5\\
&0.25$^\circ$ & 91.7 $\pm$ 0.4 & 10.4 $\pm$ 0.3 & 92.2 $\pm$ 6.0 & 5944.6 $\pm$ 0.4 & 38.5 $\pm$ 2.1\\
&\color{}{0.50$^\circ$} & 63.6 $\pm$ 0.3 & \textit{10.0 $\pm$ 0.0} & 362.5 $\pm$ 19.5 & 4711.6 $\pm$ 0.4 & 108.4 $\pm$ 5.7\\
&\color{}{1.00$^\circ$} & 64.3 $\pm$ 0.3 & \textit{10.0 $\pm$ 0.0} & \textit{395.6 $\pm$ 4.2} & 2876.0 $\pm$ 0.8 & 116.9 $\pm$ 1.4\\
&\color{}{1.50$^\circ$} & 64.8 $\pm$ 0.2 & \textit{10.0 $\pm$ 0.0} & \textit{398.0 $\pm$ 2.0} & 867.3 $\pm$ 1.4 & 116.5 $\pm$ 0.8\\
&\color{}{2.00$^\circ$} & 65.0 $\pm$ 0.2 & \textit{10.0 $\pm$ 0.0} & \textit{398.8 $\pm$ 1.2} & -2331.2 $\pm$ 0.8 & 116.3 $\pm$ 0.6\\
&\color{}{2.50$^\circ$} & 65.5 $\pm$ 0.2 & \textit{10.0 $\pm$ 0.0} & \textit{399.1 $\pm$ 0.9} & -7394.0 $\pm$ 0.4 & 115.4 $\pm$ 0.5\\
&\color{}{3.00$^\circ$} & 65.7 $\pm$ 0.2 & \textit{10.0 $\pm$ 0.0} & \textit{399.3 $\pm$ 0.6} & -14202.7 $\pm$ 0.8 & 115.1 $\pm$ 0.4\\
&\color{}{3.50$^\circ$} & 66.0 $\pm$ 0.1 & \textit{10.0 $\pm$ 0.0} & \textit{399.5 $\pm$} 0.5 & -22947.4 $\pm$ 1.4 & 114.4 $\pm$ 0.4\\
\hline
\multirow{11}{*}{\rotatebox[origin=c]{90}{\textbf{Galaxy Up}}} 
&0.00$^\circ$ & 83.4 $\pm$ 0.8 & 15.4 $\pm$ 0.5 & 123.2 $\pm$ 7.9 & 6538.4 $\pm$ 2.0 & 14.0 $\pm$ 3.0\\
&0.05$^\circ$ & 85.1 $\pm$ 1.0 & 17.9 $\pm$ 1.0 & 109.2 $\pm$ 10.8 & 6053.1 $\pm$ 1.2 & 17.2 $\pm$ 4.2\\
&\color{}{0.10$^\circ$} & 115.6 $\pm$ 1.0 & \textit{19.8 $\pm$ 0.1} & 103.1 $\pm$ 8.2 & 5241.5 $\pm$ 0.8 & 61.3 $\pm$ 2.0\\
&\color{}{0.25$^\circ$} & 116.4 $\pm$ 0.8 & \textit{19.8 $\pm$ 0.2} & 237.1 $\pm$ 13.2 & 3084.7 $\pm$ 0.4 & 103.9 $\pm$ 5.6\\
&\color{}{0.50$^\circ$} & 116.4 $\pm$ 0.5 & \textit{19.9 $\pm$ 0.1} & \textit{396.2 $\pm$ 3.5} & -263.5 $\pm$ 1.4 & 173.3 $\pm$ 1.8\\
&\color{}{1.00$^\circ$} & 118.1 $\pm$ 0.4 & \textit{19.9 $\pm$ 0.1} & \textit{399.7 $\pm$ 0.3} & -17175.9 $\pm$ 1.9 & 177.6 $\pm$ 0.7\\
&\color{}{1.50$^\circ$} & 119.2 $\pm$ 0.4 & \textit{19.9 $\pm$ 0.1} & 399.8 $\pm$ 0.2 & -51247.4 $\pm$ 0.5 & 179.4 $\pm$ 0.6\\
&\color{}{2.00$^\circ$} & 120.4 $\pm$ 0.4 & \textit{20.0 $\pm$ 0.0} & \textit{399.8 $\pm$ 0.2} & -104276.8 $\pm$ 0.5 & 181.2 $\pm$ 0.5\\
&\color{}{2.50$^\circ$} & 120.5 $\pm$ 0.4 & \textit{20.0 $\pm$ 0.0} & \textit{399.8 $\pm$ 0.2} & -177375.2 $\pm$ 0.5 & 181.4 $\pm$ 0.6\\
&\color{}{3.00$^\circ$} & 119.0 $\pm$ 0.7 & \textit{19.9 $\pm$ 0.0} & \textit{399.3 $\pm$ 0.6} & -274871.3 $\pm$ 3.0 & 179.1 $\pm$ 1.1\\
&\color{}{3.50$^\circ$} & 117.2 $\pm$ 0.5 & \textit{20.0 $\pm$ 0.0} & \textit{399.7 $\pm$ 0.3} & -393876.0 $\pm$ 0.5 & 176.5 $\pm$ 0.8\\
\hline
\end{tabular}

\caption{Recovery of an injected 21\,cm signal with a central frequency of 85\,MHz, a Signal width of 15\,MHz, and Signal Depth of 155\,mK by the REACH pipeline.
We show the evidence and accuracy of a series of recoveries with a number of models in which the antenna and ground plane are rotated westwards with respect to the model generating the data. Galaxy Down refers to an observation beginning 2019-10-01 00:00:00 and lasting three hours, during which time the galactic centre is below or near to the horizon, Galaxy Up refers to an observation beginning 2019-07-01 00:00:00 and lasting three hours, during which the galactic centre is above the horizon. log(\(\mathcal{Z}\)) is the log Bayesian evidence of our model, and RMSE refers to the root mean square error when comparing the injected mock signal to one that we generate using the posterior averages that our Gaussian model suggests. All parameters that have saturated the priors are italicised. }
\label{tab:orientation}
\end{table*}

\begin{table*}
\centering
\begin{tabular}{lllllll}
\hline
& Beam & F$_0$ (MHz) & Signal Width (MHz) & Depth (mK) & log($\mathcal{Z}$) & RMSE (mK) \\
\hline
\multirow{4}{*}{\rotatebox[origin=c]{90}{\textbf{\scriptsize Galaxy Down}}} 
&BRA & 85.9 $\pm$ 0.1 & 14.3 $\pm$ 0.1 & 146.1 $\pm$ 0.8 & 6767.4 $\pm$ 0.7 & 6.2 $\pm$ 0.4 \\
&\color{}{FRA} & 125.0 $\pm$ 0.2 & 13.8 $\pm$ 0.3 & \textit{399.9 $\pm$ 0.1} & -10033.5 $\pm$ 0.9 & 165.3 $\pm$ 1.3 \\
&\color{}{FRS} & 125.6 $\pm$ 0.1 & \textit{10.0 $\pm$ 0.0} & \textit{400.0 $\pm$ 0.0} & -799673.4 $\pm$ 0.5 & 147.8 $\pm$ 0.1 \\
&\color{}{FIS} & 125.9 $\pm$ 0.1 & \textit{10.0 $\pm$ 0.0} & \textit{400.0 $\pm$ 0.0} & -859101.5 $\pm$ 0.5 & 147.8 $\pm$ 0.1 \\
\hline
\multirow{4}{*}{\rotatebox[origin=c]{90}{\textbf{\scriptsize Galaxy Up}}} 
&BRA & 83.5 $\pm$ 0.8 & 15.4 $\pm$ 0.6 & 141.3 $\pm$ 6.7 & 3228.7 $\pm$ 1.9 & 7.8 $\pm$ 2.1 \\
&\color{}{FRA} & 106.1 $\pm$ 0.1 & 13.2 $\pm$ 0.1 & \textit{400.0 $\pm$ 0.0} & -454649.2 $\pm$ 0.5 & 131.9 $\pm$ 0.6 \\
&\color{}{FRS} & 114.7 $\pm$ 0.1 & \textit{20.0 $\pm$ 0.0} & \textit{400.0 $\pm$ 0.0} & -2035842.1 $\pm$ 2.1 & 172.6 $\pm$ 0.2 \\
&\color{}{FIS} & 114.1 $\pm$ 0.1 & \textit{20.0 $\pm$ 0.0} & \textit{400.0 $\pm$ 0.0} & -2833688.8 $\pm$ 0.5 & 171.6 $\pm$ 0.2 \\
\hline
\end{tabular}
\caption{Recovery of an injected 21\,cm signal with a central frequency of 85\,MHz, a Signal width of 15\,MHz, and Signal Depth of 155\,mK by the REACH pipeline. We show the evidence and accuracy of a series of recoveries with a number of models with the four beams we described in this paper. Galaxy Down refers to an observation beginning 2019-10-01 00:00:00 and lasting three hours, Galaxy Up refers to an observation beginning 2019-07-01 00:00:00 and lasting three hours. log($\mathcal{Z}$) is the log Bayesian evidence of our model, and RMSE refers to the root mean square error comparing the injected mock signal to one generated using posterior averages.  All parameters that have saturated the priors are italicised.}
\label{tab:beams}
\end{table*}



\bibliographystyle{mnras}
\bibliography{references} 





\bsp	
\label{lastpage}
\end{document}